




\font\bigbold=cmbx12
\font\ninerm=cmr9
\font\eightrm=cmr8
\font\sixrm=cmr6
\font\fiverm=cmr5
\font\ninebf=cmbx9
\font\eightbf=cmbx8
\font\sixbf=cmbx6
\font\fivebf=cmbx5
\font\ninei=cmmi9  \skewchar\ninei='177
\font\eighti=cmmi8  \skewchar\eighti='177
\font\sixi=cmmi6    \skewchar\sixi='177
\font\fivei=cmmi5
\font\ninesy=cmsy9 \skewchar\ninesy='60
\font\eightsy=cmsy8 \skewchar\eightsy='60
\font\sixsy=cmsy6   \skewchar\sixsy='60
\font\fivesy=cmsy5
\font\nineit=cmti9
\font\eightit=cmti8
\font\ninesl=cmsl9
\font\eightsl=cmsl8
\font\ninett=cmtt9
\font\eighttt=cmtt8
\font\tenfrak=eufm10
\font\ninefrak=eufm9
\font\eightfrak=eufm8
\font\sevenfrak=eufm7
\font\fivefrak=eufm5
\font\tenbb=msbm10
\font\ninebb=msbm9
\font\eightbb=msbm8
\font\sevenbb=msbm7
\font\fivebb=msbm5
\font\tensmc=cmcsc10


\newfam\bbfam
\textfont\bbfam=\tenbb
\scriptfont\bbfam=\sevenbb
\scriptscriptfont\bbfam=\fivebb
\def\Bbb{\fam\bbfam}

\newfam\frakfam
\textfont\frakfam=\tenfrak
\scriptfont\frakfam=\sevenfrak
\scriptscriptfont\frakfam=\fivefrak
\def\frak{\fam\frakfam}

\def\smc{\tensmc}


\def\eightpoint{%
\textfont0=\eightrm   \scriptfont0=\sixrm
\scriptscriptfont0=\fiverm  \def\rm{\fam0\eightrm}%
\textfont1=\eighti   \scriptfont1=\sixi
\scriptscriptfont1=\fivei  \def\oldstyle{\fam1\eighti}%
\textfont2=\eightsy   \scriptfont2=\sixsy
\scriptscriptfont2=\fivesy
\textfont\itfam=\eightit  \def\it{\fam\itfam\eightit}%
\textfont\slfam=\eightsl  \def\sl{\fam\slfam\eightsl}%
\textfont\ttfam=\eighttt  \def\tt{\fam\ttfam\eighttt}%
\textfont\frakfam=\eightfrak \def\frak{\fam\frakfam\eightfrak}%
\textfont\bbfam=\eightbb  \def\Bbb{\fam\bbfam\eightbb}%
\textfont\bffam=\eightbf   \scriptfont\bffam=\sixbf
\scriptscriptfont\bffam=\fivebf  \def\bf{\fam\bffam\eightbf}%
\abovedisplayskip=9pt plus 2pt minus 6pt
\belowdisplayskip=\abovedisplayskip
\abovedisplayshortskip=0pt plus 2pt
\belowdisplayshortskip=5pt plus2pt minus 3pt
\smallskipamount=2pt plus 1pt minus 1pt
\medskipamount=4pt plus 2pt minus 2pt
\bigskipamount=9pt plus4pt minus 4pt
\setbox\strutbox=\hbox{\vrule height 7pt depth 2pt width 0pt}%
\normalbaselineskip=9pt \normalbaselines
\rm}


\def\ninepoint{%
\textfont0=\ninerm   \scriptfont0=\sixrm
\scriptscriptfont0=\fiverm  \def\rm{\fam0\ninerm}%
\textfont1=\ninei   \scriptfont1=\sixi
\scriptscriptfont1=\fivei  \def\oldstyle{\fam1\ninei}%
\textfont2=\ninesy   \scriptfont2=\sixsy
\scriptscriptfont2=\fivesy
\textfont\itfam=\nineit  \def\it{\fam\itfam\nineit}%
\textfont\slfam=\ninesl  \def\sl{\fam\slfam\ninesl}%
\textfont\ttfam=\ninett  \def\tt{\fam\ttfam\ninett}%
\textfont\frakfam=\ninefrak \def\frak{\fam\frakfam\ninefrak}%
\textfont\bbfam=\ninebb  \def\Bbb{\fam\bbfam\ninebb}%
\textfont\bffam=\ninebf   \scriptfont\bffam=\sixbf
\scriptscriptfont\bffam=\fivebf  \def\bf{\fam\bffam\ninebf}%
\abovedisplayskip=10pt plus 2pt minus 6pt
\belowdisplayskip=\abovedisplayskip
\abovedisplayshortskip=0pt plus 2pt
\belowdisplayshortskip=5pt plus2pt minus 3pt
\smallskipamount=2pt plus 1pt minus 1pt
\medskipamount=4pt plus 2pt minus 2pt
\bigskipamount=10pt plus4pt minus 4pt
\setbox\strutbox=\hbox{\vrule height 7pt depth 2pt width 0pt}%
\normalbaselineskip=10pt \normalbaselines
\rm}


\def\pagewidth#1{\hsize= #1}
\def\pageheight#1{\vsize= #1}
\def\hcorrection#1{\advance\hoffset by #1}
\def\vcorrection#1{\advance\voffset by #1}

\newif\iftitlepage   \titlepagetrue               
\newtoks\titlepagefoot     \titlepagefoot={\hfil} 
\newtoks\otherpagesfoot    \otherpagesfoot={\hfil\tenrm\folio\hfil}
\footline={\iftitlepage\the\titlepagefoot\global\titlepagefalse
           \else\the\otherpagesfoot\fi}

\font\extra=cmss10 scaled \magstep0
\setbox1 = \hbox{{{\extra R}}}
\setbox2 = \hbox{{{\extra I}}}
\setbox3 = \hbox{{{\extra C}}}
\setbox4 = \hbox{{{\extra Z}}}
\setbox5 = \hbox{{{\extra N}}}

\def\RRR{{{\extra R}}\hskip-\wd1\hskip2.0 
   true pt{{\extra I}}\hskip-\wd2
\hskip-2.0 true pt\hskip\wd1}
\def\Real{\hbox{{\extra\RRR}}}    

\def\CCC{{{\extra C}}\hskip-\wd3\hskip 2.5 true pt{{\extra I}}
\hskip-\wd2\hskip-2.5 true pt\hskip\wd3}
\def\Complex{\hbox{{\extra\CCC}}}   

\def\ZZZ{{{\extra Z}}\hskip-\wd4\hskip 2.5 true pt{{\extra Z}}}
\def\Zed{\hbox{{\extra\ZZZ}}}       



\def\u{{\frak u}}
\def\s{{\frak s}}

\def\Z{{\Zed}}
\def\R{{\Real}}
\def\C{{\Complex}}

\def\pa{\partial}
\def\ket#1{|#1\rangle}
\def\bra#1{\langle#1|}


\def\abstract#1{{\parindent=30pt\narrower\noindent\ninepoint\openup
2pt #1\par}}


\newcount\notenumber\notenumber=1
\def\note#1
{\unskip\footnote{$^{\the\notenumber}$}
{\eightpoint\openup 1pt #1}
\global\advance\notenumber by 1}


\def\frac#1#2{{#1\over#2}}
\def\dfrac#1#2{{\displaystyle{#1\over#2}}}

\def\({\left(}
\def\){\right)}
\def\<{\langle}
\def\>{\rangle}

\def\pmb#1{\setbox0=\hbox{$#1$}%
   \kern-.025em\copy0\kern-\wd0
   \kern.05em\copy0\kern-\wd0
   \kern-.025em\raise.0433em\box0 }


\global\newcount\secno \global\secno=0
\global\newcount\meqno \global\meqno=1
\global\newcount\appno \global\appno=0
\newwrite\eqmac
\def\romappno{\ifcase\appno\or A\or B\or C\or D\or E\or F\or G\or H
\or I\or J\or K\or L\or M\or N\or O\or P\or Q\or R\or S\or T\or U\or
V\or W\or X\or Y\or Z\fi}
\def\eqn#1{
        \ifnum\secno>0
            \eqno(\the\secno.\the\meqno)\xdef#1{\the\secno.\the\meqno}
          \else\ifnum\appno>0
            \eqno({\rm\romappno}.\the\meqno)\xdef#1{{\rm\romappno}.\the\meqno}
          \else
            \eqno(\the\meqno)\xdef#1{\the\meqno}
          \fi
        \fi
\global\advance\meqno by1 }


\global\newcount\refno
\global\refno=1 \newwrite\reffile
\newwrite\refmac
\newlinechar=`\^^J
\def\ref#1#2{\the\refno\nref#1{#2}}
\def\nref#1#2{\xdef#1{\the\refno}
\ifnum\refno=1\immediate\openout\reffile=refs.tmp\fi
\immediate\write\reffile{
     \noexpand\item{[\noexpand#1]\ }#2\noexpand\nobreak}
     \immediate\write\refmac{\def\noexpand#1{\the\refno}}
   \global\advance\refno by1}
\def\semi{;\hfil\noexpand\break ^^J}
\def\nl{\hfil\noexpand\break ^^J}
\def\refn#1#2{\nref#1{#2}}

\def\ann#1#2#3{{\sl Ann. Phys.} {\bf {#1}} (19{#2}) #3}

\def\jmp#1#2#3{{\sl J. Math. Phys.} {\bf {#1}} (19{#2}) #3}
\def\jpA#1#2#3{{\sl J.  Phys.} {\bf A{#1}} (19{#2}) #3} 
 
\def\mplA#1#2#3{{\sl Mod.  Phys.  Lett.} {\bf A{#1}} (19{#2}) #3} 
 
\def\np#1#2#3{{\sl Nucl.  Phys.} {\bf B{#1}} (19{#2}) #3} 
\def\plA#1#2#3{{\sl Phys. Lett.} {\bf {#1}A} (19{#2}) #3}
\def\plB#1#2#3{{\sl Phys.  Lett.} {\bf {#1}B} (19{#2}) #3} 
\def\pr#1#2#3{{\sl Phys.  Rev.} {\bf {#1}} (19{#2}) #3} 
\def\prB#1#2#3{{\sl Phys.  Rev.} {\bf B{#1}} (19{#2}) #3} 
\def\prD#1#2#3{{\sl Phys.  Rev.} {\bf D{#1}} (19{#2}) #3} 
\def\prl#1#2#3{{\sl Phys.  Rev.  Lett.} {\bf #1} (19{#2}) #3}


\refn\P
{
{}For a review, see
A.M.\ Perelomov,
{\sl Physica} 4{\bf D} (1981) 1.
}
\refn\H
{F.D.M.\ Haldane,
\prl{50}{83}{1153};
\plA{93}{83}{464}.
}
\refn\TCNT
{See, for $S=\frac{1}{2}$, 
D.A.\ Tennaut, R.A.\ Cowley, S.E.\ Nagler, A.M.\ Tsvelik,
\prB{52}{95}{13368}; \nl for $S=1$, A.M.\ Tsvelik,
\prB{42}{90}{10499}.
}
\refn\WiZ
{F.\ Wilczek, A.\ Zee,
\prl{51}{83}{2250}.
}
\refn\WuZ
{
Y.S.\ Wu, A.\ Zee,
\plB{147}{84}{325}; 
}
\refn\F
{E.\ Fradkin,
\lq\lq Field Theories in Condensed Matter Systems\rq\rq, FIP 
vol.82, Addison-Wesley, New York, 1991.
}
\refn\FS
{E.\ Fradkin, M.\ Stone,
\prB{38}{88}{7215}.
}
\refn\WeZ
{X.G.\ Wen, A.\ Zee,
\prl{61}{88}{1025}.
}
\refn\HH
{F.D.M.\ Haldane,
\prl{61}{88}{1029}.
}
\refn\CHN
{
S.\ Chakravarty, B.I.\ Halperin, D.R.\ Nelson,
\prl{60}{88}{1057}.
}
\refn\SI
{G.\ Sierra,
\jpA{29}{96}{3299}.
}
\refn\SKKR
{
S.L.\ Sondhi, A.\ Karlhede, S.A.\ Kivelson, E.H.\ Rezayi,
\prB{47}{93}{16419}.
}
\refn\AP
{
W.\ Apel, Yu.A.\ Bychkov,
\prl{78}{97}{2188}.
}
\refn\VY
{
G.E.\ Volovik, V.M.\ Yakovenko,
{\it Hopf Term for a Two-Dimensional Electron Gas},
cond-mat/9703228, and references therein.
}
\refn\AH
{I.\ Affleck, F.D.M.\ Haldane,
\prB{36}{87}{5291}; \nl
I.\ Affleck,
\np{305}{88}{582}.
}
\refn\BKW
{M.J.\ Bowick, D.\ Karabali, L.C.R.\ Wijewardhana,
\np{271}{86}{417}.
}
\refn\I
{C.J.\ Isham,
{\it in} \lq\lq Relativity, Groups, and Topology\rq\rq 
$\,$ (B.S.\ DeWitt and R.\ Stora, Eds.), North-Holland, 
Amsterdam, 1984.
}
\refn\SC
{L.\ Schulman,
\pr{176}{68}{1558}.
}
\refn\LDW
{M.G.G.\ Laidlaw, C.M.\ DeWitt,
\prD{3}{71}{1375}.
}
\refn\D
{J.S.\ Dowker,
\jpA{5}{72}{936}.
}
\refn\TT
{S.\ Tanimura, I.\ Tsutsui,
\mplA{10}{95}{2607}; 
{\it Inequivalent Quantization and
Holonomy Factor from the Path-Integral Approach}, 
hep-th/9609089, to appear in {\sl Ann. Phys.}
}
\refn\Schulman
{For a review, see 
L.S. Schulman, \lq\lq Techniques and Applications of
Path Integration\rq\rq, John Willey \& Sons, New York, 1981.
}
\refn\WuZZ
{Y.S.\ Wu, A.\ Zee,
\np{272}{86}{322}.
}
\refn\BST
{A.P.\ Balachandran, A.\ Stern, G.\ Trahern,
\prD{19}{79}{2416}.
}
\refn\BMSS
{A.P.\ Balachandran, G.\ Marmo, B.S.\ Skagerstam, A.\ Stern,
\lq\lq Classical Topology and Quantum States\rq\rq,
World Scientific, Singapore, 1991.
}
\refn\M
{G.\ Mackey,
\lq\lq Induced Representation of Groups and Quantum 
Mechanics\rq\rq,
Benjamin, New York, 1969.
}
\refn\IKOS
{Y.\ Igarashi, S.\ Kitakado, H.\ Otsu, T.\ Sato,
\mplA{12}{97}{57}. 
}
\refn\Ta
{S.\ Tanimura,
\plB{340}{94}{57}.
}
\refn\LL
{N.P.\ Landsman, N.\ Linden, 
\np{365}{91}{121}.
}
\refn\MT
{D.\ McMullan, I.\ Tsutsui, 
\plB{320}{94}{287};
\ann{237}{95}{269}.
}
\refn\OK
{
Y.\ Ohnuki, S.\ Kitakado, 
\mplA{7}{92}{2477}; 
\jmp{34}{93}{2827}.
}
\refn\BT
{
R.\ Bott, L.W.\ Tu,
\lq\lq Differential Forms in Algebraic Topology\rq\rq,
GTM vol.82, Springer, New York, 1995.
}
\refn\Hu
{
D.\ Husemoller,
\lq\lq Fibre Bundles\rq\rq$\,$
(2nd edition),
Springer, New York, 1975.
}
\refn\Wh
{
G.W.\ Whitehead,
\lq\lq Elements of Homotopy Theory\rq\rq,
GTM vol.61, Springer, New York, 1978.
}


\pageheight{23cm}
\pagewidth{15cm}
\hcorrection{-1mm}
\magnification= \magstep1
\def\bsk{%
\baselineskip= 15.7pt plus 1pt minus 1pt}
\parskip=5pt plus 1pt minus 1pt
\tolerance 8000



\def\tr{{\rm Tr}\,}
\def\pa{\partial}
\def\nvec{{\bf n}}
\def\conf{{\cal Q}}
\def\calg{{\cal G}}
\def\conh{{\cal H}}
\def\CP{\C{\!\!}P^1}

\def\A{{\rm A}}
 
\def\bra{\langle}
\def\ket{\rangle}

\def\dfrac#1#2{\displaystyle\frac{#1}{#2}}



\null

{
\leftskip=105mm
\hfill\break
Revised: July 1997
\hfill\break
KEK Preprint 97-19 
\hfill\break
KU-AMP-97004
\hfill\break
hep-th/9705183
\hfill\break
\par}

\smallskip
\vfill
{\baselineskip=18pt

\centerline{\bigbold 
Quantum Mechanically Induced Hopf Term} 
\centerline{\bigbold
in the O(3) Nonlinear Sigma Model}
 
\vskip 30pt

\centerline{
\smc 
Hiroyuki Kobayashi\note
{E-mail:\quad kobayasi@tanashi.kek.jp}
\quad {\rm and} \quad 
Izumi Tsutsui\note
{E-mail:\quad tsutsui@tanashi.kek.jp}
}

\vskip 5pt

{
\baselineskip=13pt
\centerline{\it 
Institute of Particle and Nuclear Studies}
\centerline{\it 
High Energy Accelerator Research Organization (KEK),
Tanashi Branch}
\centerline{\it Midori, Tanashi, Tokyo 188, Japan}
}

\vskip 15pt

\centerline{
\smc Shogo Tanimura\note
{E-mail:\quad tanimura@kuamp.kyoto-u.ac.jp}
}

\vskip 5pt

{
\baselineskip=13pt
\centerline{\it
Department of Applied Mathematics and Physics}
\centerline{\it 
Graduate School of Engineering, Kyoto University}
\centerline{\it Kyoto 606-01, Japan}
}

\vskip 60pt

\abstract{%
{\bf Abstract.}\quad
The Hopf term in the $2 + 1$ dimensional 
$O(3)$ nonlinear sigma model, which is
known to be responsible for fractional 
spin and statistics, is re-examined from
the viewpoint of quantization ambiguity.
It is confirmed that the Hopf term can be understood as
a term induced quantum mechanically, in precisely the same
manner as the $\theta$-term in QCD.  We present a
detailed analysis of the topological aspect of the 
model based on the adjoint orbit parametrization of the
spin vectors, which is not only very useful in
handling topological (soliton and/or Hopf) numbers, but
also plays a crucial role in defining the Hopf term for 
configurations of nonvanishing soliton numbers.  
The Hopf term is seen to arise explicitly as a quantum
effect which emerges when quantizing an $S^1$ degree of freedom
hidden in the configuration space.     
}

\vfill\eject

\bsk


\noindent
{\bf 1. Introduction}
\medskip

The $O(3)$ nonlinear sigma model (NLSM)
is a very useful model, almost ubiquitous in physics,
appearing in various circumstances where the 
original $O(3)$ symmetry of the system is
spontaneously broken.  
In condensed matter physics, it describes systems
ranging from anti-ferromagnetic spin chains to
a certain class of materials which exhibit quantum Hall effects.
In particle physics, it is considered to be a prototype
of QCD, primarily because the (pure) NLSM is 
asymptotically free in $D = 1 + 1$ dimensions.
Classically, the model 
is governed by the action
$$
I =  \int d^D x \, {1 \over{2\lambda^2}}\, 
\pa_\mu {\bf n}(x)\cdot
\pa^\mu {\bf n}(x),
\eqno(1.1)
$$
where $\lambda$ is a coupling constant and
${\bf n}(x) = (n_1(x), n_2(x), n_3(x))$ is a field of
spin vectors which are subject to the constraint 
${\bf n}^2(x) = \sum_a n_a^2(x) = 1$.
Under appropriate boundary conditions, 
topological terms $I_{\rm top}$ of strength
$\theta$ (which is an angle parameter)
are available, and can be 
added to the action $I$ in (1.1).
In $1 + 1$ dimensions, the topological term is given by
[\P]  
$$
I_{\rm top} = \theta \int d^2 x \,
{1\over{8\pi}} 
\epsilon^{ij}\epsilon^{abc}\, n_a(x)\, 
\pa_i n_b(x)\, \pa_j n_c(x).
\eqno(1.2)
$$
It is known [\H] that 
the $O(3)$ NLSM can be derived as an effective low
energy theory for
the anti-ferromagnetic Heisenberg spin
chains, where the angle parameter turns out to be 
$\theta = 2\pi S$ for spin $S$ chains.
This result is 
consistent with experimental observations [\TCNT].

Meanwhile, in $D = 2 + 1$ dimensions, 
we can define the topological current,
$$
J^\mu = {1\over{8\pi}}
\epsilon^{\mu\nu\lambda}\epsilon^{abc} n_a
\pa_\nu n_b\pa_\lambda n_c,
\eqno(1.3)
$$
together with the gauge potential $A_\mu$ from 
the relation
$J^\mu = \epsilon^{\mu\nu\lambda}\pa_\nu A_\lambda$,   
and thereby furnish the topological term [\WiZ,\WuZ] 
(see also [\F]),
$$
I_{\rm top} = - \theta
\int d^3 x\, A_\mu(x)\, J^\mu(x) = \theta\, H(\nvec).
\eqno(1.4)
$$
The term $H(\nvec)$, called the Hopf (invariant) term, is nonlocal
when expressed in terms of ${\bf n}(x)$, 
but can be made local by introducing, for instance,
the $\CP$ parametrization [\WuZ] of the spin vectors.
In contrast to the
$1 + 1$ dimensional case, the 
topological term does not arise ({\it i.e.,} $\theta = 0$)
in the effective theory
of the spin chains [\FS,\WeZ,\HH], and this result is also 
experimentally supported [\CHN].  However,
it is argued [\SI] that the Hopf term does emerge
in a slightly different situation, that is, in
the spin $S$ ladders (a system consisting of $n_l$ ladders
of spin chains coupled weakly) where the angle parameter 
is given by $\theta = 2\pi n_l S$.   
A nonvanishing Hopf term has also been reported to arise  
with the value $\theta = \pi$ in the NLSM derived in describing
the quantum Hall effect at small Zeeman energies
[\SKKR,\AP,\VY].

The addition of these topological terms has major 
ramifications for the physics.  For example, 
in $1 + 1$ dimensions
it is believed [\H] that, 
unlike for the case of integral spin 
$S$, the spectrum becomes gapless for half-integral 
$S$ because the NLSM with the angle $\theta = \pi$ acquires
a fixed point with vanishing mass 
in renormalization group flow, at which 
it becomes equivalent to the $SU(2)$ 
$k = 1$ Wess-Zumino-Novikov-Witten model [\AH].
Moreover, in $2 + 1$ dimensions, it has been shown [\WiZ,\BKW] 
that fractional
spin and statistics may occur if the angle $\theta$ 
is nonvanishing.   

On the other hand, we know from the
general theory of quantization (see, {\it e.g.}, [\I]) 
that topological terms can be induced upon quantization, 
if the configuration space $\conf$ of the system is 
topologically nontrivial.
A familiar case of such configuration spaces
arises when the space $\conf$ is multiply connected 
[\SC,\LDW,\D], namely, when the fundamental group of the space 
$\pi_1(\conf)$ is nontrivial.
This case admits the appealing interpretation that induced
terms derive from the ambiguity in phase in defining
the path-integral (although a similar interpretation is
also possible for more general cases [\TT]), with 
the phase being given by a unitary 
representation of the fundamental group $\pi_1(\conf)$. 
The simplest example of this case
is the circle $\conf = S^1$ for which $\pi_1(\conf) = \Z$. 
Parametrizing the circle by an angle 
$\phi$, we find, at the quantum level, the induced 
topological term [\Schulman],
$$
I_{\rm top} = \theta
\int dt\, {1 \over{2\pi}}{{d\phi}\over{dt}}.
\eqno(1.5)
$$
The parameter $\theta$ signifies the possible 
inequivalent quantizations
(or superselection sectors), which can be determined
only by extra physical requirements, such as the amount of 
magnetic flux penetrating the circle if the system is
put in the setting of
the Aharonov-Bohm effect.  If such a requirement is not
available, the parameter is intrinsically 
indeterminable, as in the case  
of the Yang-Mills theory where the $\theta$-term
causes the strong CP problem.  

In fact, the topological terms in the $O(3)$ NLSM
are the analogue of the $\theta$-term in the Yang-Mills theory,
since they can be understood as quantum mechanically 
induced terms in the sense mentioned above.  
Thus it needs to be examined whether
or not those topological terms
pose a similar problem in the $O(3)$ NLSM, too.
Discussions from the viewpoint of 
the analogy with the Yang-Mills theory
have been given in ref.[\WuZZ] using the phase ambiguity 
interpretation
of the path-integral, which is available since 
the configuration space $\conf$
of the NLSM has $\pi_1(\conf) = \Z$.  
The aim of 
the present
paper is to provide a different argument for this on 
a firm basis, rather than using the phase ambiguity 
interpretation.
Specifically, 
we shall uncover
the fact that, in the $2 + 1$ dimensional $O(3)$ NLSM,
it is the $S^1$ degree of freedom hidden
in $\conf$ that induces the Hopf term (1.4) 
in precisely the form of the induced term (1.5) 
at the quantum level.  
In the course
of the discussion we employ the adjoint 
orbit parametrization of the spin
vector field $\nvec(x)$ in terms of an 
$SU(2)$ group valued field
$g(x)$.  This parametrization, which has been  
advocated in refs.[\BST,\BMSS], 
not only renders the Hopf term local as the $\CP$
parametrization does, but also
provides a very
convenient tool in analyzing the topological aspects of the
configurations we need to consider.
We present here a more rigorous treatment of the adjoint
orbit parametrization than previously given, 
so that the Hopf term 
can be dealt with properly in the NLSM for 
configurations of any soliton numbers.

This paper is organized as follows:
after this Introduction, in sect.2 we introduce 
the adjoint
orbit parametrization for the spin vectors in the NLSM, paying
special attention to the validity of the parametrization
in regard to the soliton numbers.  Then in sect.3
we define the soliton and Hopf
numbers to be assigned to a generic configuration
based on the adjoint orbit parametrization, 
which furnishes a coherent framework for 
dealing with topologically nontrivial 
configurations.
Sect.4 contains
an important statement about the decomposition
of a generic configuration, which allows us to extract
the $S^1$ degree of freedom from $\conf$ and induce the
Hopf term as mentioned.  Finally, sect.5 is 
devoted to discussions and outlooks.
An appendix is supplied at the end to provide a mathematical
account of the topological invariants and homotopy
groups discussed in the text.

\bigskip
\noindent
{\bf 2. Adjoint orbit parametrization}
\medskip

The parametrization of spin vectors in terms of
an adjoint orbit of $SU(2)$, or more generally 
the parametrization
of vectors taking values on a coset space $G/H$  
in terms of an adjoint orbit of the group $G$,
was introduced in ref.[\BST] in an attempt 
to formulate the NLSM as a gauge theory.  
In the $O(3)$ NLSM, we take the coset to be 
$G/H = SU(2)/U(1) \simeq S^2$, where 
$H = U(1)$ is the subgroup 
generated by, say, 
the element $T_3$
in the basis $\{T_a = {{\sigma_a}\over{2i}}; a = 1, 2, 3\}$
of the Lie algebra of $SU(2)$.
In more detail, given a field $\nvec(x)$ 
taking values on a sphere $S^2$, 
we consider a field
$g(x)$ which takes values on the group manifold $G = SU(2)$
and reproduces the spin vectors $\nvec(x)$ by the formula,
$$
g(x)\, T_3\, g^{-1}(x) = 
n(x) := n_1(x)T_1 + n_2(x)T_2 + n_3(x)T_3.
\eqno(2.1)
$$
In other words, the adjoint orbit parametrization
consists of identifying the target space $S^2$
of the spin vectors with the adjoint orbit ${\cal O}_K$
of $SU(2)$
passing through the element $K$ which we choose to be $T_3$.
In what follows we adopt the convention 
$\tr := (-2)$ times the matrix trace, which leads to  
the normalized trace $\tr(T_aT_b) = \delta_{ab}$.
Then it is seen that
the constraint satisfied by the spin vectors reads 
$\tr n^2(x) = 1$, and that this condition 
is automatically fulfilled by the parametrization (2.1).
Note, however, that the adjoint orbit parametrization
possesses redundancy, since $\nvec(x)$
is unchanged under the \lq gauge 
transformations\rq, 
$$
g(x) \longrightarrow g(x)h(x), \qquad 
\hbox{where}\quad h(x) \in H = U(1).
\eqno(2.2)
$$ 
We shall see later that the gauge
transformations are well-defined only for 
topologically trivial functions $h(x)$.

In order to discuss the class of maps that can be 
used for $g(x)$,
we now specify the class of
configurations of spin vectors $\nvec(x)$ in the NLSM
which we will be interested in. 
We first take
our spacetime to be $\R^2 \times [0, T]$.
Then the configurations of interest are those which
become a single constant vector 
at infinity in space,
$$
{\bf n}(x) = {\bf n}({\bf x}, t)
\longrightarrow {\bf n}(\infty)
\qquad \hbox{as} \quad \Vert {\bf x} \Vert \rightarrow \infty.
\eqno(2.3)
$$
The vector ${\bf n}(\infty)$ is assumed to be constant 
not only at infinity 
in space 
$\Vert {\bf x} \Vert \rightarrow \infty$ for ${\bf x} \in \R^2$ 
but also in time $t \in [0, T]$.  
Then, 
by adding a point, say, the South pole
${\bf x}_{\rm S}$ to the space $\R^2$ 
and identifying it with a sphere $S^2$,
this allows us to regard 
${\bf n}(x)$ as a map 
$S^2 \times [0, T] \rightarrow S^2$ with the value
at the South pole 
${\bf n}({\bf x}_{\rm S}, t) = {\bf n}(\infty)$ fixed for 
any $t \in [0, T]$.
The space of these maps, {\it i.e.}, 
based maps from $S^2$ to $S^2$,
is the configuration space $\conf$ of our $O(3)$ NLSM,
$$
\conf = {\rm Map}_0(S^2, S^2).
\eqno(2.4)
$$
A salient feature of the space of based maps 
is that it satisfies the following 
useful identity for homotopy groups (see Appendix),
$$
\pi_k(\conf) = \pi_{k+2}(S^2), \qquad \hbox{for} \quad k = 0, 1, 2,
\ldots.
\eqno(2.5)
$$
Thus, in particular, we have
$$
\pi_0(\conf) = \pi_{2}(S^2) = \Z,
\eqno(2.6)
$$
which states that the configuration space $\conf$  
is disconnected into \lq $n$-soliton 
sectors' $\conf_n$ labelled by 
the integers $n$ which correspond to
$\pi_{2}(S^2)$:
$$
\conf = \bigcup_{n \in \Z} \conf_n.
\eqno(2.7)
$$
We also have
$$
\pi_1(\conf) = \pi_{3}(S^2) = \Z,
\eqno(2.8)
$$
which implies that 
each soliton sector is multiply connected
$\pi_1(\conf_n) = \Z$, hence providing 
the basis for the existence of 
the Hopf term (1.4).
The basic motivation for employing 
the adjoint orbit parametrization is 
to gain a better control over this topological structure
of the configuration space by making use of the group property.

Returning to the field $g(x)$, 
the first question we need to address is the
very existence of the map $S^2 \times [0, T] 
\rightarrow SU(2) \simeq S^3$ that will be assigned to
$g(x)$ if it is to 
correspond under (2.1) to
the map $S^2 \times [0, T] \rightarrow S^2$ given by
$\nvec(x)$.  The answer
is negative, however, because if there was such a map then 
there would also be a map $S^2 \rightarrow S^3$
(from the target $S^2$ to the target $S^3$ at a fixed time), 
but this is impossible since the $S^3$, 
regarded as a $U(1)$ principal bundle over the base space
$S^2$, is known to be nontrivial in general.
(For more detail, see Appendix.)  
An obvious wayout
is to trivialize the bundle, that is, first 
remove the South pole and consider the map
$D^2 \times [0, T] \rightarrow S^2$ for 
$\nvec(x)$, where $D^2$ is 
a two dimensional
disc of unit radius which is identified with $S^2 - \{ 
{\bf x}_{\rm S} \}$, and then find a map
$D^2 \times [0, T] \rightarrow S^3$ for $g(x)$
related to $\nvec(x)$ under (2.1).  When this is done,
the map $g(x)$ must reproduce the constant spin vector
$\nvec(\infty)$
on the boundary of the disc $\pa D^2 = S^1$, that is,
$$
g(x)\, T_3\, g^{-1}(x) = n(\infty), \qquad
\hbox{for} \quad x \in \pa D^2 \times [0, T].
\eqno(2.9)
$$
If we let $g(\infty)$ be 
a representative fixed element fulfilling (2.9), 
then, due to the ambiguity under (2.2),
we find that $g(x)$ takes the following form
on the boundary,
$$
g(x) = g(\infty) h(x), \qquad h(x) \in H = U(1),
\qquad \hbox{for} \quad x \in \pa D^2 \times [0, T],
\eqno(2.10)
$$
where $h(x)$ is an arbitrary function of space and time.  
At this point we remark
that the function $h(x)$ in (2.10), which
characterizes the field
$g(x)$ on the boundary $\pa D^2$, can be nontrivial as a map
$S^1 \rightarrow U(1)$, whereas
those functions $h(x)$ used for gauge transformations (2.2)
must be trivial because they define a $U(1)$ bundle over
the disc $D^2$, which is trivial since $D^2$ is contractible.

To sum up the foregoing argument
in terms of spaces, consider first the space of maps for 
$g(x)$,
$$
\calg = {\rm Map}'(D^2, SU(2)),
\eqno(2.11)
$$
where the prime indicates that
maps in $\calg$ fulfill the condition (2.10).
Let now
$$
\conh = {\rm Map}(D^2, U(1)),
\eqno(2.12)
$$
be the space of functions $h(x)$ used for gauge transformations
(2.2).  Then, the adjoint orbit parametrization amounts to 
identifying the configuration space $\conf$ 
of the spin vectors $\nvec(x)$ given in (2.4) 
with the quotient space,
$$
\conf = \calg/\conh,
\eqno(2.13)
$$
where $\calg/\conh$ is the orbit space of $\calg$ under
$\conh$, {\it i.e.}, the set of equivalent classes under
gauge transformations (2.2).

An important point to note is 
that the space $\calg$, which is characterized by 
the boundary value $g(\infty)$ on $\pa D^2$, 
does not in general form
a group.  In fact, 
multiplication may not be well-defined in $\calg$ 
since it does not necessarily 
respect the boundary condition (2.10), or 
the inverse $g^{-1}$ may not exist in $\calg$ since 
$g^{-1}$ may not obey (2.10) even if $g$ does. 
There are, however, two cases of $\calg$
worth noticing.  One is the case 
where $g(\infty) \in H$ on $\pa D^2$. 
In this case, the space $\calg$, which we 
denote by $\calg^{+}$, 
does form a group,
enjoying the property $g^+\, {g^+}' \in \calg^{+}$
and $(g^+)^{-1} \in \calg^{+}$ for 
any $g^+$, ${g^+}'$ in $\calg^{+}$.
A remarkable property of the space $\calg^{+}$ is that
it preserves any space $\calg$ under right-multiplication,
$$
g\, g^+ \in \calg \qquad \hbox{for} 
\quad g \in \calg, \quad g^+ \in \calg^+.
\eqno(2.14)
$$
The other space, which we denote by $\calg^{-}$, 
occurs when $g(\infty) \in e^{\pi T_2} H $
on $\pa D^2$.
Although the space $\calg^{-}$ 
does not itself form a group, it satisfies
$(g^-)^{-1} \in \calg^{-}$ for $g^-$ in $\calg^{-}$ and
$$
g^- {g^-}'\in \calg^{+}, \qquad
g^- g^+ \in \calg^{-}, \qquad
g^+ g^- \in \calg^{-},
\eqno(2.15)
$$
where we have used 
$T_3 e^{\pi T_2} = - e^{\pi T_2} T_3$.  For 
this reason we may assign `$\pm$' (`even' and 
`odd') parity to $\calg^{\pm}$,
respectively.  These spaces $\calg^{\pm}$ are
actually homeomorphic to each other under 
the parity operation\note{
Notice the difference from the spin flip operation
$\nvec(x) \rightarrow - \nvec(x)$ implemented by
$g \rightarrow g\, e^{\pi T_2}$.}
$$
g \longrightarrow e^{\pi T_2}\, g.
\eqno(2.16)
$$ 
We also mention that, in addition to these two spaces, 
there is a space $\calg^{\rm c}$, which too forms 
a group,
consisting of 
configurations which become constant 
$g(x) = g_{\rm c}$ on $\pa D^2$.
If we define the space of 
all configurations of the type (2.10) with (not a fixed
but) any $g(\infty) \in SU(2)$,
$$
\widetilde{\calg}
 = \bigcup_{g(\infty)\in SU(2)} \calg,
\eqno(2.17)
$$
then we find that in $\widetilde{\calg}$ the subspace
$\calg^{\rm c}$ intersects 
with the other subspaces $\calg^{+}$ and $\calg^{-}$ at
$g_{\rm c} = 1$ and 
$g_{\rm c} = e^{\pi T_2}$ modulo $H = U(1)$
as shown in Fig.1.

\topinsert
\vskip 1.4cm

\let\picnaturalsize=N
\def\picsize{5cm}
\def\picfilename{fig1.epsf}
\ifx\nopictures Y\else{\ifx\epsfloaded Y\else\input epsf \fi
\global\let\epsfloaded=Y
\ifx\picnaturalsize N\epsfxsize \picsize\fi 
\hskip 3.25cm\epsfbox{\picfilename}}\fi
                
\abstract{%
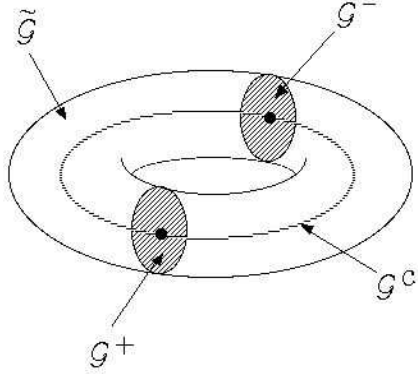
{\bf Figure 1.} A schematic picture of the spaces $\calg^+$, 
$\calg^-$ and $\calg^{\rm c}$ in $\widetilde{\calg}$ 
(which is represented 
by a doughnut $D^2 \times S^1$).  Each of the two intersection 
points in $\calg^{\rm c}$ actually consists of a circle $S^1$; 
one with $\calg^+$ is given by $H$ and the other with 
$\calg^-$ by $e^{\pi T_2} H$.
}
\endinsert

Let us now consider under what circumstances 
multiplication is well-defined in 
the total space $\widetilde{\calg}$.
To this end, note first that the condition
$g \in \widetilde{\calg}$ 
is equivalent to 
$$
g^{-1} dg \in \u(1), \qquad \hbox{for} 
\quad x \in \pa D^2 \times [\,0, T\,],
\eqno(2.18)
$$
where $\u(1)$ is the Lie algebra of $H = U(1)$
generated by $T_3$.  Thus, for the product
$g\,g'$ formed from $g$, $g' \in \widetilde{\calg}$
to belong again to $\widetilde{\calg}
$ we must have
$(g\,g')^{-1} d (g\,g') \in \u(1)$, 
but this holds if and only if we have on the boundary,
$$
g'(\infty)^{-1}\, (g^{-1}dg) \, g'(\infty) \in \u(1),
\eqno(2.19)
$$ 
where $g'(\infty)$ is the constant fixed element that 
characterizes $g'$ on the boundary $\pa D^2$.
Parametrize  
the element $g'(\infty)$ by the
Euler angles $(\rho, \varphi, \chi) \in [0,2\pi] \times
[0,\pi] \times [0,4\pi]$ as
$$
g'(\infty) = e^{\rho T_3}\, e^{\varphi T_2}\,
e^{\chi T_3}.
\eqno(2.20)
$$
Since $g^{-1}dg \in \u(1)$ on $\pa D^2$, we can 
write $g^{-1}dg = \gamma_3 T_3$ with a 1-form $\gamma_3$.
Then, a straightforward computation yields
$$
g'(\infty)^{-1}\, (g^{-1}dg) \, g'(\infty) 
= \gamma_3\, \bigl[ \cos\varphi\, T_3 - \sin\varphi\, 
(\cos\chi\, T_1 - \sin\chi\, T_2) \bigr].
\eqno(2.21)
$$
We thus find that, for (2.19) to be valid, we need
either (i) $\gamma_3 = 0$, (ii) $\varphi = 0$ or 
(iii) $\varphi = \pi$.  
Conversely, (2.19) holds if any of (i) -- (iii) holds.
Observe that (i) is the case where  
$g \in \calg^{\rm c}$, whereas (ii) and
(iii) imply $g' \in \calg^+$ and $g' \in \calg^-$,
respectively.  

In conclusion, multiplication in $\widetilde{\calg}$
is well-defined
if and only if (at least) one of the two elements 
belong to those subspaces in the manner 
stated above.  Note also that the space $\calg$
is not necessarily preserved under multiplication,
except for case (ii) which is 
the right-multiplication by $\calg^+$
mentioned earlier.  For instance, gauge 
transformations (2.2) preserve
the space since
they belong to case (ii), whereas 
global $SU(2)$ left-transformations,
$$
g(x) \longrightarrow g_{\rm c}\,g(x), \qquad \hbox{where}
\quad g_{\rm c} \in SU(2),
\eqno(2.22)
$$ 
do not, since they belong to case (i).  The latter
transformations lead to 
$O(3)$ rigid rotations of the vectors ${\bf n}(x)$
through (2.1), which are the symmetry transformations
of the action (1.1).

\bigskip
\noindent
{\bf 3. Topological (soliton and Hopf) numbers} 
\medskip

The advantage of using the adjoint orbit parametrization
becomes manifest when we deal with the soliton number  
which is the charge of the topological current (1.3),
$$
Q({\bf n}) = \int_{S^2}d^2x\, J^0(x).
\eqno(3.1)
$$
The soliton number $Q({\bf n})$ 
is actually the wrapping number of
the map $S^2 \rightarrow S^2$ given by 
${\bf n}(x)$ at any fixed time,
and hence takes an integer
related to the homotopy group (2.6).
On the other hand, the soliton number is introduced 
also to $g(x)$ by 
$$
Q(g) = - {1\over{4\pi}} \int_{\pa D^2}\, \tr T_3 
(g^{-1}(x)dg(x)).
\eqno(3.2)
$$
Indeed, changing 
the domain of the integral in (3.1)
to $D^2$ by removing the South pole ${\bf x}_{\rm S}$
from $S^2$ (which does not change the outcome
of the integral) and using the relation (2.1),
we can easily confirm that the expression
$Q({\bf n})$ in (3.1) reduces to $Q(g)$ in (3.2).
It then follows from (2.10) that the soliton
number (3.2) is nothing but the 
winding number of the map 
$\pa D^2 \simeq S^1 \rightarrow H = U(1)$ given by 
$h(x)$ on the boundary $\pa D^2$.
In other words, the piece $h(x)$ in
$g(x)$ in (2.10) which corresponds 
to a given spin vector field $\nvec(x)$ is chosen 
such that
its winding number be equal to the soliton
number $Q({\bf n})$ and thereby relates
the two integers, $\pi_1(U(1)) = \Z$ and $\pi_2(S^2) = \Z$. 
As we remarked earlier, gauge transformations (2.2) are
trivial on the boundary $\pa D^2$ 
as a map $S^1 \rightarrow U(1)$ and 
cannot change the winding number, that is,
the soliton number $Q(g)$ is gauge invariant, as required.
We also note that the zero soliton sector $\conf_0$
is the only sector where the $U(1)$ bundle 
over $\pa D^2 \simeq S^1$ is trivial and hence
$h(x)$ can be chosen to be constant on the boundary, 
that is, the sector where 
$g(x)$ is well-defined as a map $S^2 \times [0, T] 
\rightarrow SU(2) \simeq S^3$.

Before we proceed, 
let us recapitulate the above result 
using the spaces mentioned
earlier: first, the fact that
any field $g(x)$ can be characterized
by the soliton number shows that,
like the configuration 
space (2.7), the space $\calg$
consists of $n$-soliton sectors $\calg_n$ which are 
disconnected each other,
$$
\calg = \bigcup_{n \in \Z} \calg_n,
\eqno(3.3)
$$
with the integer $n$ being 
the element of $\pi_0(\calg) = \Z$.
Second, the fact that gauge transformations 
preserve each sector $\calg_n$ means that 
$$
\conf_n = \calg_n/\conh.
\eqno(3.4)
$$  
 
We have learned in sect.2 that multiplication in 
$\widetilde{\calg}$ is well-defined only for 
the cases (i) -- (iii).  It is interesting
to note that the soliton number becomes additive
or subtractive
depending on the cases, that is,
$$
Q(gg') = \pm \, Q(g) + Q(g'),
\eqno(3.5)
$$
where the `$+$' sign holds for case (ii) 
where $g' \in \calg^+$, whereas the `$-$'
holds for case (iii) where $g' \in \calg^-$.
Indeed, from (3.2) we obtain
$$
Q(gg') - Q(g') = 
- {1\over{4\pi}} \int_{\pa D^2}\, \tr (g' T_3 g'^{-1}) 
(g^{-1}dg),
\eqno(3.6)
$$
but the r.h.s.~turns out to be just $\pm Q(g)$,
since from (2.20) we have
$
g'(x) T_3 g'(x)^{-1} 
= g'(\infty) T_3 g'(\infty)^{-1} = \pm T_3,
$
where `$+$' holds for (ii) 
and `$-$' for (iii). For case (i) the r.h.s.~of 
(3.6) vanishes, which shows that the soliton number
$Q(g)$ is invariant under global $SU(2)$
left-transformations (2.22).  Thus, 
in particular, the soliton number
is preserved 
$Q(e^{\pi T_2} g) = Q(g)$ under the parity operation (2.16).

{}From the property (3.5) it follows that 
the inverse $g^{-1}$ of an even element 
$g \in \calg^+$ has the opposite soliton charge 
$Q(g^{-1}) = - Q(g)$, and therefore the
configuration $g^{-1}$ may be called `anti-soliton'
with respect to $g$.
On the other hand, the inverse $g^{-1}$ of 
an odd element $g \in \calg^-$ 
has the same soliton charge $Q(g^{-1}) = Q(g)$, 
and hence in this case we may take the transpose
${}^{\rm t}g$ of $g$ to get an anti-soliton, 
on account of $Q({}^{\rm t}g) = - Q(g)$.  
We would need to go through a more
involved procedure to deduce these results
based on the additive/subtractive 
property of the soliton number, 
if we used the original spin vectors $\nvec(x)$.

Let us next consider how the Hopf number  
in (1.4) reads in the adjoint orbit parametrization. For this, 
we first restrict ourselves to 
the 0-soliton sector $\conf_0$, and consider a 
boundary condition
periodic in time such that ${\bf n}(x)$ becomes a constant vector 
${\bf n}(\infty)$ both at
$t = 0$ and $T$,
$$
{\bf n}({\bf x}, T)  = {\bf n}({\bf x}, 0)  
= {\bf n}(\infty).
\eqno(3.7)
$$
This allows us to
identify all points on the space $S^2$ at 
$t = 0$ and reduce them to a single point (similar reduction can 
be done for the space $S^2$ at $t = T$),   
whereby regard the spacetime under consideration as $S^3$.
Hence, in this case ${\bf n}(x)$ provides
a map $S^3 \rightarrow S^2$, which is known to be classified
by the Hopf number which characterizes
the homotopy group (2.8).  Corresponding to (3.7), 
we may consider the
boundary condition for the field $g(x)$ as
$$
g({\bf x}, T) = g({\bf x}, 0)\, h_{\rm c} 
= g(\infty), 
\eqno(3.8)
$$
where $g(\infty)$ is the element
specified in (2.10) and $h_{\rm c} \in H$ 
is an arbitrary constant element.
Then, a similar 
reasoning employed for ${\bf n}(x)$ allows us to  
regard that the spacetime for the map $g(x)$ is $S^3$
and, accordingly, $g(x)$ provides a map
$S^3 \rightarrow SU(2)$.
Upon using the relation (2.1), we find
after a little algebra
that the Hopf term $H(\nvec)$ in (1.4) becomes local
[\BMSS] (see also Appendix),
$$
H(g) = {1\over{48\pi^2}} \int_{S^3}\, \tr (g^{-1}(x) dg(x))^3.
\eqno(3.9)
$$
Since this gives 
the degree of mapping $S^3 \rightarrow SU(2)$,
the identity
$H(g) = H({\bf n})$ implies
the relation between the two integers, 
$\pi_3(SU(2)) = \Z$ and $\pi_3(S^2) = \Z$. 

We now proceed to define the Hopf number
to configurations which have nonvanishing soliton numbers.
To this end, let us impose
the following boundary condition in time, 
$$
g({\bf x}, T) = g({\bf x}, 0)\, h_{\rm c},
\eqno(3.10)
$$
namely, we leave out
the constancy condition in space imposed in (3.8), as it
can be implemented only in the 0-soliton sector. 
We then consider the configuration,
$$
\bar g(x) := g(\infty)\, 
g^{-1}({\bf x}, 0) \, g({\bf x}, t),
\eqno(3.11)
$$
which still lies in the space $\calg$ if $g \in \calg$.
Moreover, we observe that $\bar g(x)$ defined in (3.11)
has the soliton number zero $\bar g \in \calg_0$
since it reduces to the identity
element at $t = 0$, and that it fulfills 
the boundary condition (3.8)
with $\bar g(\infty) = 1$ (which implies 
$\bar g \in \calg^+$).  
It is thus admissible to assign
the Hopf number to a configuration $g \in \calg_n$ 
of any soliton number $n$ first by converting it to
$\bar g \in \calg_0$ and then evaluating the value
of the Hopf invariant $H(\bar g)$.

To illustrate our point, 
let us introduce
the coordinates
$(\alpha, \beta) \in [0, 2\pi] \times [0, \pi]$
on the disc $D^2$ by
${\bf x} =
\bigl({{\beta}\over{\pi}} \cos\alpha,
 {{\beta}\over{\pi}} \sin\alpha\bigr)$
with the boundary $\pa D^2$ being identified with points having
$\beta = \pi$, and consider
the familiar 1-soliton 
configuration which evolves dynamically
with respect to the collective coordinate $\phi(t)$:
$$
{\bf n}_{1}(x) = \bigl(\cos(\alpha + \phi(t))\sin\beta,\,
\sin(\alpha + \phi(t))\sin\beta,\, \cos\beta\bigr).
\eqno(3.12)
$$
The corresponding expression in the adjoint orbit
parametrization is
$$
g_{1}(x) = e^{\phi(t)T_3}\, g_1({\bf x}),
\eqno(3.13)
$$
where $g_1({\bf x})$ stands for the 
static 1-soliton (Skyrmion) configuration,
$$
g_1({\bf x}) = e^{\alpha T_3}\, e^{\beta T_2}\,
e^{- \alpha T_3}. 
\eqno(3.14)
$$
Note that the 1-soliton configuration $g_1({\bf x})$
belongs to
the parity odd subspace $\calg^-$ 
and, since $e^{\phi(t)T_3} \in \calg^+$, 
the configuration $g_{1}(x)$ in
(3.13) also belongs to $\calg^-$.  We also find that
$g_{1}(x)$ possesses a
unit soliton number and further fulfills 
(3.10) if $\phi(t)$ satisfies a proper periodic boundary
condition, say, 
$$
\phi(0) = 0 \qquad \hbox{and} \qquad \phi(T) = 2m\pi,
\qquad \hbox{with} \quad m \in \Z.
\eqno(3.15)
$$
Then, according to (3.11) we consider
$$
\bar g_{\rm 1}(x) 
= e^{\pi T_2}\, 
g_1^{-1}({\bf x})\, e^{\phi(t)T_3}\, g_1({\bf x}).
\eqno(3.16)
$$
In order to evaluate the Hopf number for 
this $\bar g_{1}(x)$,
we first change the spacetime from $S^3$ to 
$D^2 \times [0, T]$, which can be done by 
proceeding conversely to what we did before.
Although this procedure is in principle 
unnecessary for a 0-soliton
configuration such as $\bar g_{1}(x)$, 
this allows us to evaluate the integral of 
the Hopf number for $\bar g_{1}(x)$ from 
its constituent pieces appearing in the decomposition
(3.16) which are defined on $D^2 \times [0, T]$.  
Indeed, a direct computation gives
$$
H(\bar g_{1}) 
= {1\over{2\pi}}
\int_0^T dt\, {{d\phi}\over{dt}} \times Q(g_1) = m,
\eqno(3.17)
$$
which shows that the zero soliton 
configuration (3.16) just constructed
possesses the Hopf number $m$.

Returning to a generic configuration $\bar g(x)$,
we observe that, unlike the soliton
number,
the Hopf number so defined always
enjoys the additive property, 
$$
H(\bar g \bar{g}') = H(\bar g) + H(\bar{g}'), 
\eqno(3.18)
$$
once multiplication
is well-defined for any 
0-soliton configurations $\bar g$, $\bar{g}'$.
This can be readily
confirmed by substituting $\bar g \bar{g}'$ 
directly in (3.9), which yields
$$
H(\bar g \bar{g}') - H(\bar g) - H(\bar{g}') = 
- {1\over{16\pi^2}} \int_{S^3}\, d\, 
\tr ({\bar g}^{-1}d{\bar g})(d\bar{g}'\,\bar{g}'^{-1}).
\eqno(3.19)
$$
Then we find that the r.h.s.~vanishes identically
since $S^3$ has no boundary, establishing
the additivity of the Hopf number.
With the help of this additive property, 
we can check that the Hopf number is 
invariant under 
the gauge transformation (2.2) which, for 
our $\bar g(x)$ in (3.11), amounts to 
$$
\bar g(x) \longrightarrow 
g(\infty)\, h^{-1}({\bf x}, 0)\, g^{-1}(\infty)\,
\bar g(x) \, h(x).
\eqno(3.20)
$$
Note that 
the product configuration appearing 
in the r.h.s.~belongs to $\calg$ if $g \in \calg$, 
and that it has the soliton number zero
because gauge transformations are trivial on the
boundary, {\it i.e.}, $Q(h) = 0$.
Then, from 
$H(h^{-1}({\bf x}, 0)) = H(h(x)) = 0$ we find
$$
H(\bar g) 
\longrightarrow 
H(g(\infty)\, h^{-1}({\bf x}, 0)\, g^{-1}(\infty)
\, \bar g(x) \, h(x)) = 
H(\bar g),
\eqno(3.21)
$$
as announced.

\bigskip
\noindent
{\bf 4. Hopf term as a quantum effect}
\medskip

Having assigned the Hopf number as well as the soliton number
to a generic configuration, we now 
come to the point to
show that the Hopf term in the $2 + 1$ dimensional
NLSM can be regarded as a quantum mechanically 
induced term.  More explicitly, we show
that
in the configuration space $\conf$
there exists a degree of freedom 
represented by an angle, and that this  
$S^1$ degree of freedom 
is responsible for inducing the Hopf term at the
quantum level.

{}For the sake of brevity, we restrict our attention again to
the 0-soliton sector $\conf_0$ where we can use $g(x)$ 
rather than $\bar g(x)$ for the Hopf number.  We do not
lose generality by this, because
any configuration
of nonvanishing soliton number can always be brought to
a 0-soliton configuration by the conversion procedure (3.11).  
Now, suppose that the given configuration 
$g(x) \in \calg$ has the Hopf number $m$.
We then consider 
$$
\hat g(x) := g(x)\, g_{\rm st}^{-1}(x), 
\eqno(4.1)
$$
where $g_{\rm st}(x) = e^{-\pi T_2}\,\bar g_1(x)$ 
is the \lq standard'
configuration possessing
the Hopf number $m$ and the soliton number 0,
constructed with the help of $\bar g_1(x)$ 
in (3.16).
Note that since 
$g_{\rm st}^{-1}$
in (4.1) belongs to the even parity subspace $\calg^+$,
we have $\hat g(x) \in \calg$ by (2.14).
Making use of the additive property (3.5) 
of the soliton number valid for case (ii),
we obtain
$$
Q(\hat g(x)) = Q(g(x)) + Q(g_{\rm st}^{-1}(x)) = 0,
\eqno(4.2)
$$
showing that the configuration 
$\hat g(x)$ has zero soliton number.
Moreover, using the additivity of the Hopf number
(3.18) we find that $\hat g(x)$ has zero
Hopf number, too.
We therefore conclude, by solving (4.1) in favor of $g(x)$, 
that any 0-soliton configuration $g(x)$ with the Hopf number
$m$ can be decomposed into the form,
$$
g(x) = \hat g(x) \, g_{\rm st}(x).
\eqno(4.3)
$$
In words,
any 0-soliton configuration
can be written as a product of
the specific 0-soliton configuration $g_{\rm st}(x)$ 
having Hopf number $m$, and
the `trivial' configuration 
$\hat g(x)$ having vanishing 
soliton and Hopf numbers.    
The point to be stressed is that,
in the generic configuration of vanishing soliton number,
the dynamical degree of freedom that governs
the Hopf number is the angle variable $\phi(t)$
in $g_{\rm st}(x)$.  As stated above, this
is also valid to any
configuration $g(x)$ 
of nonvanishing soliton number,
on account of the conversion procedure (3.11).

This result needs to be examined more carefully,
since one needs to pay attention to the distinction between
the two spaces of maps, the true configuration space
$\conf$ in (2.4) and the space of maps $\calg$ in (2.11),
which are related by (2.13).  
However, from the fact 
that gauge transformations
(2.2) are right-transformations which do not change the
soliton number nor the Hopf number (because these
topological 
numbers are gauge invariant) we learn that 
the degrees of freedom associated with the
gauge transformations represented by $\conh$
reside entirely in the last piece
$\hat g(x)$ in the decomposition (4.3).
Stated in terms of spaces, (4.3) implies 
the decomposition $\calg_0 = S^1 \times \widehat \calg_0$, 
where $\widehat \calg_0$ is the space of 
trivial configurations $\hat g(x)$ for which
$\pi_1(\widehat \calg_0) = 0$.  Thus, combining the 
gauge invariance of the $S^1$ degrees of freedom together
with (3.4) for $n = 0$, we find
$$
\conf_0 = S^1 \times \bigl( \widehat \calg_0 / \conh \bigr).
\eqno(4.4)
$$
Since a similar decomposition is valid to any
soliton sector $\conf_n$, we conclude that
$$
\conf = S^1 \times \bigl( \widehat \calg / \conh \bigr),
\eqno(4.5)
$$
where $\widehat \calg = \bigcup_{n} \widehat \calg_n$.
In short, 
the configuration space $\conf$ admits a decomposition
in such a way that the gauge invariant
angle degree of freedom represented 
by the circle $S^1$ is factored out.

Now, let us consider the quantization of the NLSM 
starting with
the classical action (1.1).  This, however, is a formidable
task to do in practice,  
because the configuration space $\conf$ is nowhere close 
in structure to a Euclidean space, for which 
we know how to quantize by, {\it e.g.},
the canonical quantization programme.
However, we also know the consequence of quantization
on relatively simple but nontrivial spaces, such as 
coset spaces $G/H$ by, {\it e.g.}, Mackey's
quantization procedure [\M].  In particular, when the 
configuration space is a circle $S^1$, we know from
Mackey's quantization procedure 
or Schulman's analysis of 
path-integral quantization [\Schulman] that, at the quantum level,
an induced term arises in the form (1.5) with 
parameter $\theta$ which labels the 
superselection sectors of the quantum theory.

This implies that, when
we focus on the quantization of the $S^1$ degree of freedom
in the configuration space $\conf$
given by the angle $\phi(t)$,
we obtain precisely the induced term (1.5) as a quantum effect.
Then, from the relation (3.17) we obtain
$$
I_{\rm top} = 
\theta\, H(g_{\rm st}) = \theta\, H(\bar g) 
= {{\theta}\over{48\pi^2}} \int_{S^3}\, \tr (\bar g^{-1}(x) 
d\bar g(x))^3.
\eqno(4.6)
$$
This shows that the Hopf term $H(\bar g)$ --- 
rather than $H(g)$ which is ill-defined for configurations
of nonvanishing
soliton numbers --- can be acquired as an induced term,
purely from quantization.

\bigskip
\noindent
{\bf 5. Discussions and outlooks}
\medskip

We have seen in this paper that the Hopf term
in the $O(3)$ NLSM in $2 + 1$ dimensions can be 
induced quantum mechanically as a manifestation
of the (inevitable) ambiguity 
in quantizing the $S^1$ degree of freedom
of the configuration space.  The tool we employed 
throughout our argument is
the adjoint orbit parametrization (2.1), which not only 
renders the Hopf term local, as does the commonly used 
$\CP$
parametrization, but also is very useful
in dealing with configurations of nonvanishing 
topological (soliton and/or Hopf) numbers.
In particular, it was shown that
the additivity of those topological numbers, (3.5) and (3.18),
which is available in the adjoint orbit parametrization,
allows decomposing any given configuration into constituent
configurations of certain topological numbers.
This enables us to extract a specific
degree of freedom, which in our case is $S^1$,
in the true configuration space $\conf$.  It should be
noted that the $S^1$ degree of freedom is 
extracted from a generic configuration, 
rather than being introduced
as a collective coordinate around the 1-soliton
configuration (3.12) as often done in, {\it e.g.}, 
refs.[\BKW,\IKOS]. 
It should also be stressed that 
the Hopf number can be assigned to sectors of nonvanishing
soliton numbers only after the conversion to the vanishing
soliton sector is performed.  To our knowledge, this conversion
has not been discussed before, possibly because 
with spin vectors $\nvec(x)$ the procedure would become
quite involved.

Our investigation on the quantum mechanical aspect
of the topological structure of the space $\conf$
in the NLSM is a continuation of a 
study in the Abelian sigma model in $1 + 1$ dimensions [\Ta]. 
In both of the models, 
we extracted an $S^1$ degree of freedom from the configuration 
space after decomposition, 
and found that the quantum effect associated with the $S^1$ 
does induce the topological term as anticipated 
in the path-integral quantization [\WuZ,\WuZZ].
This result reinforces the view that the induced
terms appearing in these models are intrincically
of quantum origin, just like the $\theta$-term in QCD.
We need to note, however, that
the investigation carried out thus far is still
a primitive one, since it 
is concerned only with the fundamental group $\pi_1(\conf)$
of the configuration space, and in this sense it would be
natural to expect more to occur
at the quantum level from the topological structure 
associated with higher homotopy groups such as,
in the case of the NLSM,
$\pi_2(\conf) = \Z_2$, $\pi_3(\conf) = \Z_2$
and so forth.
Indeed, it is known [\LL,\MT] that there arises
an induced term given by the Dirac
monopole potential 
when one quantizes 
on a sphere $\conf = S^2$ for which $\pi_2(\conf) = \Z$, 
and further, 
that a similar
phenomenon occurs on $S^n$ in general (see also [\OK]).  
Although it is not yet clear how to realize such 
a phenomenon in the NLSM, we expect that our direct
method of extracting 
the crucial degree(s) of freedom from the space $\conf$ 
is more fruitful than
the phase ambiguity consideration in 
the path-integral [\WuZ,\WuZZ,\TT].

In this respect, it might be worth mentioning the 
same NLSM placed in $1 + 1$ dimensions.  
In this case, if one  
proceeds analogously to the $2 + 1$ dimensional case
assuming the boundary condition on the spin
vector field similarly to (2.3) {\it etc.}, 
the configuration space will be
$\conf = {\rm Map}_0(S^1, S^2)$.
Then, using an argument similar to that used for the
$2 + 1$ dimensional case, one obtains
$\pi_0(\conf) = \pi_{1}(S^2) = 0$, which shows that
the configuration space is connected and hence has no
distinct soliton sectors.  On the other hand, one also
observes that $\pi_1(\conf) = \pi_{2}(S^2) = \Z$, which
indicates that 
the space $\conf$ is multiply connected, 
implying that there exist configurations possessing
topological (winding) numbers dynamically, bearing 
the term (1.2).
One further notices that
$\pi_2(\conf) = \pi_{3}(S^2) = \Z$, whose 
quantum mechanical consequence may 
appear as an induced potential of the Dirac
monopole as mentioned earlier.  
The adjoint orbit parametrization
will continue to be useful in studying this case, 
although one needs
to take care of the maps $g(x)$ anew in regard to
the topology of the
configuration space, as we did for the $2 + 1$ dimensional
case in the present paper.    

\bigskip
\noindent
{\bf Acknowledgements:}
I.T. is grateful to L.~Feh\'er for helpful 
discussions on homotopy groups
and M.~Barton for valuable comments, and
S.T. wishes to thank T.~Iwai and Y.~Uwano for 
their encouragement and useful discussions.
This work is supported in part by 
the Grant-in-Aid for Scientific
Research from the Ministry of Education, Science and Culture
(No.09740199).

\bigskip

\noindent
{\bf Appendix}
\medskip

The purpose of this Appendix is to
put in order some of the basic mathematical ingredients 
of the $2 + 1$ dimensional $O(3)$ NLSM, 
{\it i.e.}, mappings and topological invariants 
(which are at times handled loosely in the text) 
based on more rigorous definitions (see, {\it e.g.}, [\BT]).
To this end, we use different notations from 
those used in the text in order to  
distinguish the various mappings strictly,
while in the main text we preferred to use simpler notations.
We also give a proof 
of the identity (2.5) for homotopy groups.

\medskip
\noindent
{\it Soliton number:} 
We begin by defining the soliton number in the NLSM.
By the canonical embedding $ S^2 \hookrightarrow \R^3 $
a point of $ S^2 $ can be identified 
with a unit vector $ \nvec \in \R^3 $.
The 2-form
$$
	\omega 
   := \frac{1}{8 \pi} \nvec \cdot ( d \nvec \times d \nvec ) 
	\eqno(\A.1)
$$
on $ S^2 $ is closed and has the integral 
$ \int_{S^2} \omega = 1 $,
and hence is a generator of the de Rham cohomology
$H^2_{\rm DR}(S^2)$.  With the coordinates
$ \nvec = ( \sin \theta \cos \phi, \sin \theta \sin \phi, 
\cos \theta ) $ it reads
$ \omega = \dfrac{1}{4 \pi} \sin \theta d \theta \wedge d \phi $.
The $ O(3) $ NLSM is described by
the field variable $ \varphi : M \to S^2 $,
where $ M $ is the base spacetime, 
which in this paper is assumed to 
be $M = S^2 \times [0, T]$.
Restriction to a particular 
time $ t \in [0, T]$ provides
the configuration $ \varphi_t : S^2 \times \{ t \} \to S^2 $
on a time-slice in $ M $, where the former $ S^2 $ represents
the base space and the latter $ S^2 $ is the target space.
Then the 
pullback of $ \omega $ by $ \varphi $ gives 
a closed 2-form $J := \varphi^* \omega $ on $ M $,
which is called topological current whose
integration
over the fixed time-slice yields an integer,
$$
	Q ( \varphi_t ) := \int_{ S^2 } \varphi_t^* \omega.
	\eqno(\A.2)
$$
In fact, $ Q ( \varphi_t ) $ counts the wrapping number of
the mapping $ \varphi_t : S^2 \to S^2 $.  It therefore
is independent of time $t$ giving the integers related to
the homotopy group $ \pi_2 ( S^2 ) = \Z $, and hence 
can be used to define 
the `soliton number' to the field $ \varphi $
as a topological charge.

\medskip
\noindent
{\it Hopf invariant:} 
There is another functional to characterize the homotopy group  
$ \pi_3 ( S^2 ) = \Z $, that is, the Hopf invariant.  
It is defined as follows.  Assume that 
the field 
$ \varphi : S^2 \times [ \, 0, T \, ] \to S^2 $ 
becomes constant 
at end times, {\it i.e.}, both 
$ \varphi_{0} : S^2 \times \{ 0 \} \to S^2 $ and
$ \varphi_{T} : S^2 \times \{ T \} \to S^2 $ 
are constant mappings.  Then 
it can be regarded as a mapping $ \varphi : S^3 \to S^2 $.
Take a 1-form $ A $ on $ S^3 $ which satisfies the equation,
$$
	d A = J = \varphi^* \omega.
	\eqno(\A.3)
$$
Note that a solution for $ A $ exists, since 
$ H_{\rm DR}^2 ( S^3 ) = 0 $ implies that
the closed 2-form $ \varphi^* \omega $ is exact on $S^3$.
The solution is of course not unique, because
if $ A $ is a solution, $ A + df $ is also a solution for 
any function $ f $ on $ S^3 $.
However, the Hopf invariant (which we also refer to as the
Hopf number) defined by
$$
	H( \varphi ) := - \int_{S^3} A \wedge J,
	\eqno(\A.4)
$$
is free from the arbitrariness thanks to
the closedness (current conservation) $ dJ = 0 $.

\medskip
\noindent
{\it Hopf fibration by the adjoint orbit parametrization:} 
The topological nature of the Hopf invariant 
becomes transparent by considering 
the Hopf fibration.  
Let $ T_a = \dfrac{\sigma_a}{2i} $  ($ a =1$, 2, 3)  
be a basis of the Lie algebra $\s\u(2)$, with which 
the inner product 
$ \bra X, Y \ket := - 2 $ tr $ ( X Y ) $ for 
$X$, $Y \in \s\u(2)$ is equipped. 
Then, using the 
isomorphism between $\s\u(2)$ and $ \R^3 $ provided by
the identification of $ X = \sum_a X_a T_a \in \s\u(2)$ 
and $ {\bf X} = ( X_1, X_2, X_3 ) \in \R^3 $,
we furnish the Hopf fibration $ p : S^3 \to S^2 $
with fibre $S^1$ 
using the mapping 
given by the adjoint action
on a fixed element, say, $T_3$ of $ \s\u(2) $ as
$p : g \mapsto {\rm Ad}(g) T_3 = g T_3 g^{-1}$.
The fibre is identified with the subgroup $ U(1) $
of $ SU(2) $ generated by $ T_3 $.

If we pullback 
the 2-form $ \omega $ on $ S^2 $ to
$ SU(2) \simeq S^3 $ by $ p $, then again by
$ H_{\rm DR}^2( S^3 ) = 0 $ the closed 2-form 
$ p^* \omega $ is exact on the $S^3$,
and therefore there exists a 1-form 
$ \rho $ on $ S^3 $ such that $ p^* \omega = d \rho $.
If we put $ g^{-1} dg = \sum_{a} \gamma_a T_a $, we find
$$
	p^* \omega
	= \frac{1}{8 \pi}
	\bra g T_3 g^{-1}, 
	[ \, d(g T_3 g^{-1}), d(g T_3 g^{-1}) \, ] \ket
	= \frac{1}{4 \pi} \gamma_1 \wedge \gamma_2
	= - \frac{1}{4 \pi} d \gamma_3,
	\eqno(\A.5)
$$
which shows that the solution is given by the 1-form
$ \rho 
= - \dfrac{1}{4 \pi} \gamma_3 
= - \dfrac{1}{4 \pi} \bra T_3, g^{-1} dg \ket $
up to an exact 1-form.

Suppose that, 
given $ \varphi : S^3 \to S^2 $, there exists 
a mapping 
$ \tilde{\varphi} : S^3 \to S^3 $ satisfying 
the following commutative diagram,
$$
	\matrix{
	    &                          & S^3             \cr
	    & \tilde{\varphi} \nearrow & 
	    \Bigl\downarrow \, p \Bigr. \cr
	S^3 & \longrightarrow          & S^2             \cr
	    &          \varphi         &                 \cr
	}
	\eqno(\A.6)
$$
Then we see that the Hopf invariant gives just the
degree of mapping $\tilde{\varphi}: 
S^3 \rightarrow S^3$ (from the base
space $S^3$ to the target $S^3$).
Indeed, since $ p \circ \tilde{\varphi} = \varphi $,
we have
$$
	d A 
	= J 
	= \varphi^* \omega 
	= ( p \circ \tilde{\varphi} )^* \omega
	=     \tilde{\varphi}^* ( p^* \omega )
	=     \tilde{\varphi}^* ( d \rho )
	= d ( \tilde{\varphi}^*     \rho ),
	\eqno(\A.7)
$$
which shows that 
$ 
A = 
\tilde{\varphi}^* \rho 
= - \dfrac{1}{4 \pi} \tilde{\varphi}^* \gamma_3 
$
is a solution for (\A.3).  When this is combined with
$ 
J =
\dfrac{1}{4 \pi} \tilde{\varphi}^* ( \gamma_1 \wedge \gamma_2 ) 
$
which is obtained from (\A.5), we find that
the Hopf invariant (\A.4) reads
$$
	\eqalign{
	H( \varphi ) 
	& = 
	\frac{1}{(4 \pi)^2} \int_{S^3} 
	\tilde{\varphi}^* ( \gamma_1 \wedge \gamma_2 \wedge \gamma_3 )
	\cr
	& = 
	\frac{1}{6 (4 \pi)^2} \int_{S^3} 
	\tilde{\varphi}^* 
	\bra g^{-1} dg, [ \, g^{-1} dg, g^{-1} dg \, ] \ket.
	}
	\eqno(\A.8)
$$
The fact that the 3-form $ \omega^{(3)} $, which is
defined by writing
$ H = \int_{S^3} \tilde{\varphi}^* \omega^{(3)} $,
gives a generator of $ H_{\rm DR}^3 ( S^3 )$
can be seen explicitly by 
introducing 
the coordinates on $ SU(2) $ by 
$ ( \phi, \theta, \psi ) \in 
 [\, 0,  2 \pi \, ] \times
 [ \, 0,  \pi \, ] \times 
 [ \, 0, 4 \pi \, ] 
$ and put the element $g \in SU(2)$ as
$g = e^{ \phi   T_3 }e^{ \theta T_2 } e^{ \psi   T_3 }$.
This leads to
$$
\eqalign{
g^{-1} dg 
&= 
( \sin \psi d \theta - \cos \psi \sin \theta d \phi ) T_1 
\cr
&\qquad +
( \cos \psi d \theta + \sin \psi \sin \theta d \phi ) T_2 +
        ( \cos \theta d \phi + d \psi ) T_3,
        }
        \eqno(\A.9)
$$
and hence
$$
\int_{SU(2)} \omega^{(3)} 
= 
\int_{SU(2)} 
\sin \theta d \theta \wedge d \phi \wedge d \psi
	= 1,
	\eqno(\A.10)
$$
as required.  We recall that
the spectral sequence associated with the exact sequence
$ S^1 \to S^3 \to S^2 $ provides the exact sequence
of homotopy groups,
$$
	\cdots \to
	\pi_3( S^1 ) \to
	\pi_3( S^3 ) \to
	\pi_3( S^2 ) \to
	\pi_2( S^1 ) \to 
	\cdots
	\eqno(\A.11)
$$
Since $ \pi_3( S^1 ) = \pi_2( S^1 ) = 0 $,
we have $ \pi_3( S^3 ) = \pi_3( S^2 ) = \Z $.
The correspondence between $ \tilde{\varphi} $ and $ \varphi $
realizes this isomorphism between 
$ \pi_3( S^3 )$ and $ \pi_3( S^2 )$.

\medskip
\noindent
{\it Existence of the map $\tilde{\varphi}$:}
The foregoing statement for the Hopf
invariant rests entirely on the assumption that
the mapping $ \tilde{\varphi} $ exists, which is nontrivial.
Before answering to the question 
whether such a mapping really exists or not,
let us first consider the static version of the problem, 
that is, the question whether the map 
$ \tilde{\sigma} : S^2 \to S^3 $ 
satisfying the commutative diagram,
$$
	\matrix{
	    &                         & S^3             \cr
	    & \tilde{\sigma} \nearrow & 
	    \Bigl\downarrow \, p \Bigr. \cr
	S^2 & \longrightarrow         & S^2             \cr
	    & \sigma                  &                 \cr
	}
	\eqno(\A.12)
$$
exists to a given static configuration 
$ \sigma : S^2 \to S^2 $.
Below we shall 
show that the necessary and sufficient condition for
the existence of $ \tilde{\sigma} $ is that 
the soliton number vanishes $ Q( \sigma ) = 0 $.
The necessity is easy; 
if we assume $ \tilde{\sigma} $ to exist, then
by definition we find
$$
	Q( \sigma )
	=
	\int_{S^2} \sigma^* \omega
	=
	\int_{S^2} ( p \circ \tilde{\sigma} )^* \omega
	=
	\int_{S^2} \tilde{\sigma}^* ( p^* \omega )
	=
	\int_{S^2} \tilde{\sigma}^* ( d \rho )
	=
	\int_{S^2} d ( \tilde{\sigma}^* \rho )
	=
	0.
	\eqno(\A.13)
$$
To show the sufficiency,
let $ D^2 $ denote a two-dimensional disk of unit radius
$ D^2 := \{ {\bf x} = (x,y) \in \R^2 | x^2 + y^2 \le 1 \} $, and
consider its one-point compactification by
$$
	\matrix{
	c : & D^2 ( \subset \R^2 ) & \to & S^2 ( \subset \R^3 ),
	\cr
	&
	\left(
		\dfrac{\theta}{\pi} \cos \phi, 
		\dfrac{\theta}{\pi} \sin \phi  
	\right)
	& \mapsto &
	 ( \sin \theta \cos \phi, 
	    \sin \theta \sin \phi, 
	    \cos \theta ),
	\cr
	}
	\eqno(\A.14)
$$
where $ 0 \le \theta \le \pi $ and $ 0 \le \phi < 2 \pi $.
The mapping $ c $ brings
all points on the boundary $ \partial D^2 $ 
to the South pole ${\bf x}_{\rm S} = (0,0,-1) \in S^2 $.
Then,
since $ D^2 $ is contractible, for 
any $ \sigma : S^2 \to S^2 $
there exists $ \hat{\sigma} : D^2 \to S^3 $ which satisfies
the commutative diagram,
$$
	\matrix{
	    & \hat{\sigma}          &      \cr
	D^2 & \longrightarrow       & S^3  \cr
	c \, \Bigl\downarrow \Bigr. &      &
	\Bigl\downarrow \, p \Bigr.        \cr
	S^2 & \longrightarrow       & S^2  \cr
	    & \sigma                &      \cr
	}
	\eqno(\A.15)
$$
The mapping $ \hat{\sigma} $ is not uniquely
determined, since it can be replaced by the 
gauge transformed one 
$ \hat{\sigma}' = \hat{\sigma} \cdot h $
with $ h : D^2 \to U(1) $.  However, 
this ambiguity does not affect the soliton number
$Q(\sigma)$ given by
$$
	Q( \sigma )
	=
	\int_{D^2} d ( \hat{\sigma}^* \rho )
	=
	\int_{\partial D^2} \hat{\sigma}^* \rho
	=
	- \frac{1}{4 \pi} 
	\int_{\partial D^2} 
        \hat{\sigma}^* \bra T_3, g^{-1} dg \ket.
	\eqno(\A.16)
$$
Clearly, the soliton number 
$ Q(\sigma) $ in (A.16)
counts the degree of the restricted mapping
$ \hat{\sigma}| : 
\partial D^2 \simeq S^1 \to p^{-1}
( \sigma ( {\bf x}_{\rm S} )) \simeq S^1 $.
Thus, if 
$ Q(\sigma) = 0 $ then $ \hat{\sigma} $ admits 
a gauge transformation to
a mapping $ \hat{\sigma}' : D^2 \to S^3 $ 
which is constant along the boundary $ \partial D^2 $.
Hence in this case we can shrink $ \partial D^2 $
to a point to get a mapping $ \tilde{\sigma} $
satisfying the diagram,
$$
	\matrix{
	    & \hat{\sigma}'         &      \cr
	D^2 & \longrightarrow       & S^3  \cr
	c \, \Bigl\downarrow \Bigr. & 
	\tilde{\sigma} \nearrow     & 
	\Bigl\downarrow \, p \Bigr.        \cr
	S^2 & \longrightarrow       & S^2  \cr
	    & \sigma                &      \cr
	}
	\eqno(\A.17)
$$
as we wanted.

Now, to answer the question of the existence of
the mapping $ \tilde{\varphi} $
satisfying (\A.6), we recall that 
$ \varphi : S^3 \to S^2 $ is regarded as
$ \varphi : S^2 \times [ \, 0, T \, ] \to S^2 $ 
which takes constant values
at end times of the period $ [ \, 0, T \, ] $.
But for this to be the case the mapping
$ \varphi_0 : S^2 \to S^2 $ must have 0-soliton number and
hence $ \varphi_t $ also has 
$ Q( \varphi_t ) = 0 $ for any $ t \in [ \, 0, T \, ]$.
Consequently, the above argument shows that
there indeed exists a map $ \tilde{\varphi}_t $ and hence
$ \tilde{\varphi} $ satisfying (\A.6), 
if $Q(\varphi_t ) = 0$.

On the other hand, this also shows that, if
$ Q( \varphi_t ) \ne 0 $ for $ \varphi : 
S^2 \times [ \, 0, T \, ] \to S^2 $,
then there cannot exist a mapping
$ \tilde{\varphi} : S^2 \times 
[ \, 0, T \, ] \to S^3 $ satisfying
$ p \circ \tilde{\varphi} = \varphi $.
However, since 
the mapping $\hat \sigma$ in diagram (A.15) always
exists, the mapping $\hat \varphi$ satisfying
$$
        \matrix{
                                  & \hat{\varphi}   &      \cr
        D^2 \times [ \, 0, T \, ] & \longrightarrow & S^3  \cr
        c \, \Bigl\downarrow \Bigr. &               &
        \Bigl\downarrow \, p \Bigr.                        \cr
        S^2 \times [ \, 0, T \, ] & \longrightarrow & S^2  \cr
            & \varphi                &                     \cr
        }
        \eqno(\A.18)
$$
is also guaranteed to exist.
The configuration $g(x)$ used for the adjoint
parametrization to describe a generic 
(nonvanishing) soliton sector
in the text is this mapping $\hat{\varphi}$.

\medskip
\noindent
{\it Configurations of nontrivial topological numbers:}
Configurations
possessing a nonvanishing soliton or 
Hopf number may be furnished explicitly as follows.
Take some integer $n \in \Z$ and consider the mapping,
$$
	\matrix{
	\sigma_n : & S^2 & \to & S^2 
	\cr
	&
	( \sin \theta \cos \phi, 
	  \sin \theta \sin \phi, 
	  \cos \theta ) 
	& \mapsto &
	( \sin \theta \cos n \phi, 
	  \sin \theta \sin n \phi, 
	  \cos \theta ).
	\cr
	}
	\eqno(\A.19)
$$
The lift $ \hat{\sigma}_n: D^2 \to S^3 $ defined in (\A.15) 
is given by $\hat{\sigma}_n({\bf x}) = e^{   n \phi T_3 }
        e^{   \theta T_2 }
        e^{ - n \phi T_3 }$.
Then it is readily confirmed from (A.16)
that the configuration has soliton number $n$,
$$
	Q( \sigma_n )
	=
	- \frac{1}{4 \pi}
	\int_{\partial D^2} \hat{\sigma}_n^* \gamma_3
	=
	- \frac{1}{4 \pi}
	\int_{S^1} ( - 2 n d \phi )
	=
	n.
	\eqno(\A.20)
$$

We may next use this configuration to construct
a configuration which has a nonvanishing Hopf number.
This can be achieved by taking the mapping
$
\hat{\tau}_m :
       [ \, 0, T \, ] \to S^1 ( \subset S^3 )
$ 
given by
$
\hat{\tau}_m(t) = e^{ 2\pi m (t/T) \, T_3 }
$ 
and thereby construct the mapping 
$
\hat{\varphi} :
       D^2 \times [ \, 0, T \, ] \to S^3
$
by
$
\hat{\varphi}({\bf x},t) := \hat{\sigma}_n^{-1}({\bf x})
\hat{\tau}_m(t) \hat{\sigma}_{n}({\bf x}).
$
It is then straightforward to show that
$$
	\bra 
	  \hat{\varphi}^{-1} d \hat{\varphi},
	[ \, 
	\hat{\varphi}^{-1} d \hat{\varphi},
	\hat{\varphi}^{-1} d \hat{\varphi} 
	\, ]
	\ket
	= 
	24\pi \, m \, {{dt}\over T} \,
	\bra
	T_3,
	d ( d \hat{\sigma}_n \hat{\sigma}_n^{-1} )
	\ket,
	\eqno(\A.21)
$$
which leads to the Hopf number,
$$
	H( \varphi ) 
	=
	\frac{4\pi}{(4 \pi)^2}
	\int_0^{T} \!\! m \, {{dt}\over T}
	\int_{\partial D^2}
	\bra
	T_3,
	d \hat{\sigma}_n \hat{\sigma}_n^{-1}
	\ket
	=
	m \, n.
	\eqno(\A.22)
$$
The case $ n = 1 $ provides a configuration having 
a unit Hopf number $ H = m $ (with $Q = 0$), 
and this is the standard
configuration $g_{\rm st}(x)$ used in the text.

\medskip
\noindent
{\it Proof of the homotopy identity (2.5):}
{}Finally, for completeness we provide a proof\note{
We owe this to L. Feh\'er.}
of the identity (2.5) for homotopy groups. First we 
note that for compact spaces $X$, $Y$ and $Z$, 
we have the homeomorphism [\Hu] between the spaces of
based maps,
$$
{\rm Map}_0\bigl(X, \,{\rm Map}_0(Y, Z)\bigr)
= {\rm Map}_0(X \wedge Y, Z),
\eqno(\A.23)
$$
where the wedge stands for the smash product of the
two spaces.  Take $X = S^k$, $Y = S^n$ and $Z = S^m$
with nonnegative intergers $k$, $n$ and $m$,
and recall the relation 
$S^k \wedge S^n = S^{k + n}$ 
valid for spheres [\Wh].
Then it follows that
$$
{\rm Map}_0\bigl(S^k, \,{\rm Map}_0(S^n, S^m)\bigr)
= {\rm Map}_0(S^{k + n}, S^m).
\eqno(\A.24)
$$
The identity (2.5) is obtained by considering the
homotopy classes for the pointed space,
$$
\pi_k\bigl({\rm Map}_0(S^n, S^m)\bigr) = \pi_{k + n}(S^m),
\eqno(\A.25)
$$
and setting $n = m = 2$.

\vfill\eject

  \vfill\eject\immediate\closeout\reffile
  \centerline{{\bf References}}\bigskip\frenchspacing%
  \input refs.tmp\vfill\eject\nonfrenchspacing

\bye